\documentclass{acm_proc_article-sp}

\usepackage{multirow}
\usepackage{graphicx}
\usepackage{longtable}
\usepackage[latin1]{inputenc}     
\usepackage[T1]{fontenc}          
\usepackage[francais]{babel} 

\begin{document}

\title{Mutation Analysis for Security}

%
%
%
%
%

\numberofauthors{4} 
%
\author{
%
%
\alignauthor ENNAHBAOUI Mohammed\\
       \affaddr{Laboratory of Mathematics, Computing and Application}\\
       \affaddr{University of Mohammed V AGDAL, Faculty of Sciences Rabat}\\
       \affaddr{P.O. Box 1014 RP. Rabat, Morocco}\\
       \email{ennahbaoui.mohamed@gmail.com}
\alignauthor EL HAJJI Said\\
       \affaddr{Laboratory of Mathematics, Computing and Application }\\
       \affaddr{University of Mohammed V AGDAL, Faculty of Sciences Rabat}\\
       \affaddr{P.O. Box 1014 RP. Rabat, Morocco}\\
       \email{elhajji.said@gmail.com}
}


\maketitle
\begin{abstract}
Security has become, nowadays, a major concern for the organizations as the majority of  its applications are exposed to Internet, which increases the threats of security considerably. Thus, the solution is to improve tools and mechanisms to strengthen the protection of applications against attacks and ensure the different security objectives. Among solutions we will talking about, in this paper, there is Mutation Analysis which is a technique of test  that evaluates the quality of software tests and their ability to detect errors, It also compares the criteria and test generation strategies. In this study we will use the Mutation Analysis as a mean to qualify the penetration tests, and then, apply this technique in the security mechanisms and exactly on the mechanisms of access control. At the end we will propose a method for the elimination of hidden mechanisms for access control that will allow the access control policy to evolve.
\end{abstract}


\section{Introduction}
Web applications become a need for companies and organizations as main communication tools and commercial windows. Yet, this need is limited by security threats that increase significantly every year. Among security threats cited, there are security flaws which the number of announced ones continues to rise, according to statistics of the Common Vulnerabilities and Exposures dictionary (CVE) that counts more than 5000 new security flaws every year and until May 2012, there are more than 2000 new security attacks detected.\\
Thus, the solution to this problematic is to develop security tools and mechanisms, and improve the existing ones to strengthen the protection of applications against attacks, taking into consideration the different security objectives (integrity, confidentiality, availability, authentication and non-repudiation). Among security techniques there are Access Control that ensures users access to resources respecting a well defined security policy. This technique reinforces the confidentiality, integrity as well as availability of information, hence its importance in the security in general.\\
One of the mechanisms that we will propose for the security problems mentioned in this study is the Mutation Analysis, considered as a testing method that evaluates quality software testing and its ability to detect errors, also compares the criteria and test generation strategies. This technique is detailed in the second part of the first section.\\
In this paper, we will talk in general, about the Mutation Analysis applied in the computer security with two different ways. The first one is the qualification of penetration test  detailed in the last part of the first section and the second one is ensuring conformity between the code which implants security and the Policy of Access Control.\\
We will use this study of Mutation Analysis application to propose approaches and solutions to test access control mechanisms and interpret their robustness, as well as to detect hidden ones. Then, we will give a concrete method for the evolution of security policy without any problems that will prevent modification of access control mechanisms. This method is detailed in the last part of the second section.\\
The first part of first section explains the basic principles and provides a brief overview of some popular tools and approaches in relation with securing applications. In parallel, the first part in the second section gives introduction and definitions for Access Control and its mechanisms.

\section{The Mutation Analysis and Penetration Tests}
\subsection{The steps for securing the applications}
There are several approaches that aim the application level security, among these approaches are those that seek to directly secure the code,using static analysis and proof techniques. These techniques vary from parsing to the formal verification code and much more others. In our article, we will present some examples of formal approaches like the Proof-Carrying Code \cite{Besson:2006:PCC:1226601.1226603} and the "Splint" tool \cite{Evans:2002:ISU:624647.626359}.\\
A second type of approaches is to secure modeling in an abstract way based on the requirements and security policy that we will establish using the Or-Bac model presented.

\subsubsection{Proof and static analysis applied to security}
As part of formal approaches we will mention two ones.\\
The first approach is the PCC (Proof-Carrying Code) introduced by Necula and Lee as a new oncoming in which the producer provides, at the same time, a compiled program and validity proof (certificate)of this software. \\
Before the client implements the program, the proof will be automatically verified by the user that possesses the tool (an installed base called Trusted Computing Base or TCB) which gives if the proof is valid for the provided program or not.\\
The proof can concern any security policy defined by the consumer, and any property of the program such as security memory, as well as limiting access to the disk, or bank transfers verification.\\
\begin{figure}[h!]                                                       %
\centering                                                               %
\includegraphics[width=5cm,height=2.5cm]{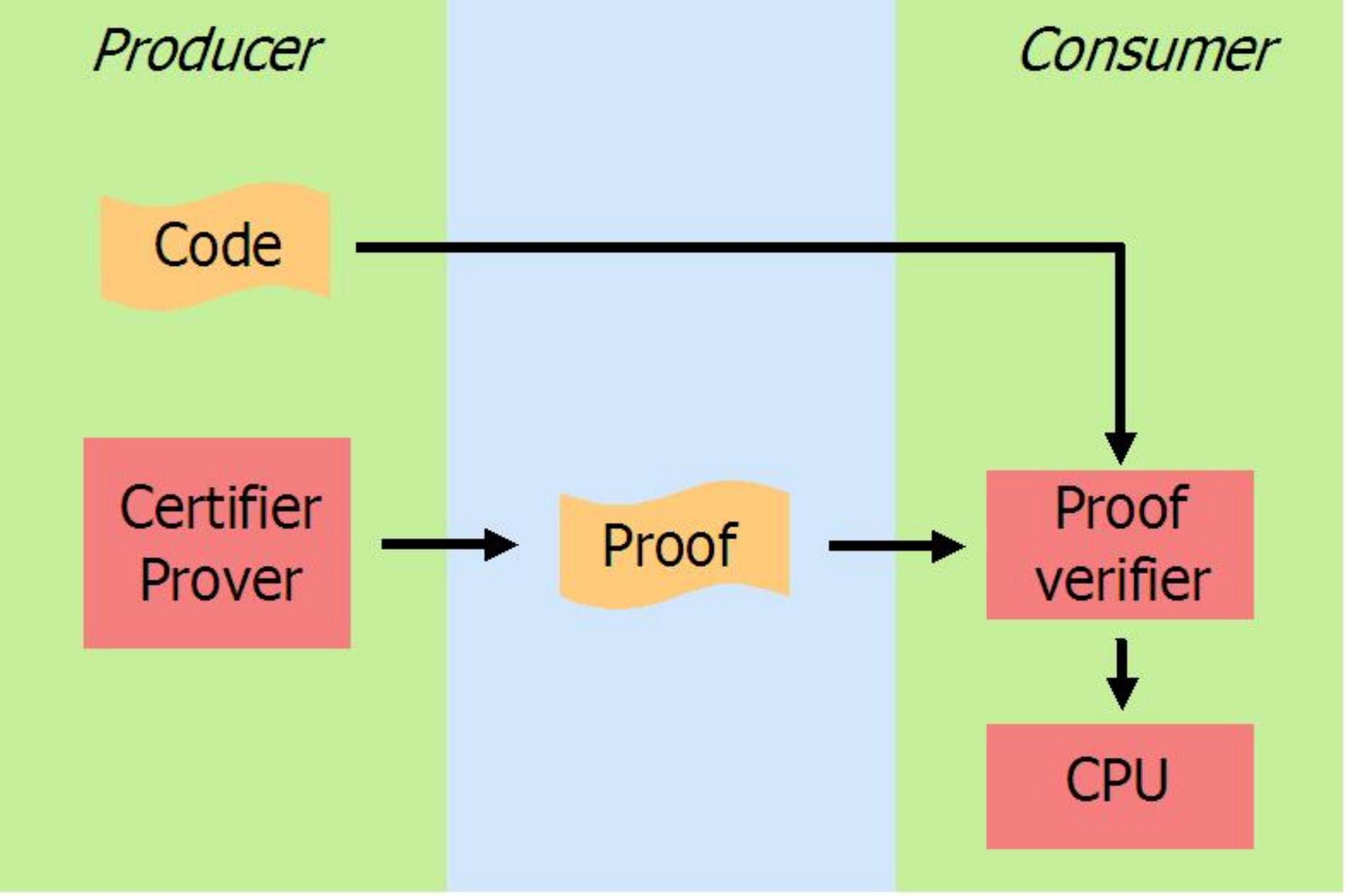}                       %
\caption{The working of PCC}                                             %
\label{The working of PCC}                                               %
\end{figure}                                                             %

It proves easy to check the certificate than creating it; hence, the use of PCC is promising because it can adapt to consumer needs, requires less blind confidence and does not involve time overhead at execution, as example, the case of mobile phone.\\

The second approach is the Static Analysis used to identify security flaws in the code. Most tools that use this approach are generally based on an abstract compilation, data streams and interpretation techniques to calculate an approximation of program behavior.\\
One of these tools is Splint, it has the capacity to identify places in the code which may represent security flaws.
The mentioned tool is based on annotations made in the source code. These annotations are the assumptions of developer concerning the progress of the execution (non-null parameter, dynamic memory allocation ...). Therefore, Splint examines the code and ensures that these assumptions are verified by Static Analysis. It helps to list the places in the code that are vulnerable to buffer overflow attacks; hence the use of such approaches is recommended during the development cycle but they are complex to implement and difficult to reach, which justify that they are little used in practice. For example, Splint produces many negatives faults. This is unfortunately the case with most tools for detecting buffer overflow as ARCHER, BOON, Splint, UNO, ...\\
The second disadvantage of these tools is that they fail to distinguish between the vulnerable code and the corrected code (detection alert even after the correction of the problem location).

\subsubsection{Modeling of requirements and security policies}
The conception approach of security models is interesting because it identifies new types of threats other than ones on code level.\\
In principle, a security policy is based on a mix between security needs and logic job (This answer is generally given as result of consultation between security experts and business experts).\\
Indeed, there are a lot of security models, among them the model (Organization Based Access Control)\cite{Anas,DBLP:conf/acsac/CuppensM03} that was developed as part of the RNRT MP6 (Models and Security Policies of Information Systems and Communication in Health and Social).\\
This model is one of the most interesting ones because it addresses the modeling of security policies in an abstract way at the organizational level independently of implantation, which means that the established OrBAC policy (called concrete) is derived from organizational policies. This conceptual approach makes all retrieved policy in the OrBAC model, reproducible and scalable. Indeed, it doesn't require any readjustment at the concrete level that could introduce an incoherence which is difficult to recover. Consequently, everything is done at the organizational level.\\
First of all, it's worth to mention what is an Access Control model.\\
Access Control is the fact to grant privileges to subjects in order to perform actions on objects.
\begin{itemize}
 \item Subjects can be users, which means individuals and the process they are using.
 \item Objects are also called resources, they can be system files, relations in a database, printers, ...
 \item The possible actions in an information system are "read", "write", "execute", or in a database "select", "update"...
\end{itemize}

The OrBAC objective is to enable modelling of variety of security policies based on the context of the organization. To achieve this purpose, and to reduce the complexity of managing access rights, the OrBAC model relays on four main principles:
\begin{itemize}
 \item The organization is the main entity of the model. It is an organized group of active entities; which means, subjects that plays some roles. Note that a group of subjects isn't necessarily considered as an organization. In other words, each subject playing a role in the organization, corresponds to a certain agreement between subjects to form an organization. In particular, organization can be structured into several sub-organizations, each one with its own security policy. As the structure of real organizations can be made up of departments, services, ...  A project or a work group can be modelled by an organization.
 \item Two levels of abstraction (OrBAC Interactions):
 \begin{itemize}
  \item a concrete level: subject, action, object
  \item an abstract level: role, activity, view
 \end{itemize}
 \item The opportunity to express permissions, prohibitions, and obligations.
 \item The opportunity to express contexts.
\end{itemize}
Thus, in addition to have a security policy independent of its implementation, OrBAC has other advantages, it can express permissions, prohibitions and obligations, it also takes in consideration the contexts, hierarchies and delegation.\\
The introduction of a level also allows the structuring of entities as seen in the following diagram:
\newpage                                                                 %
\begin{figure}[h!]                                                       %
\centering                                                               %
\includegraphics[width=7cm,height=3.5cm]{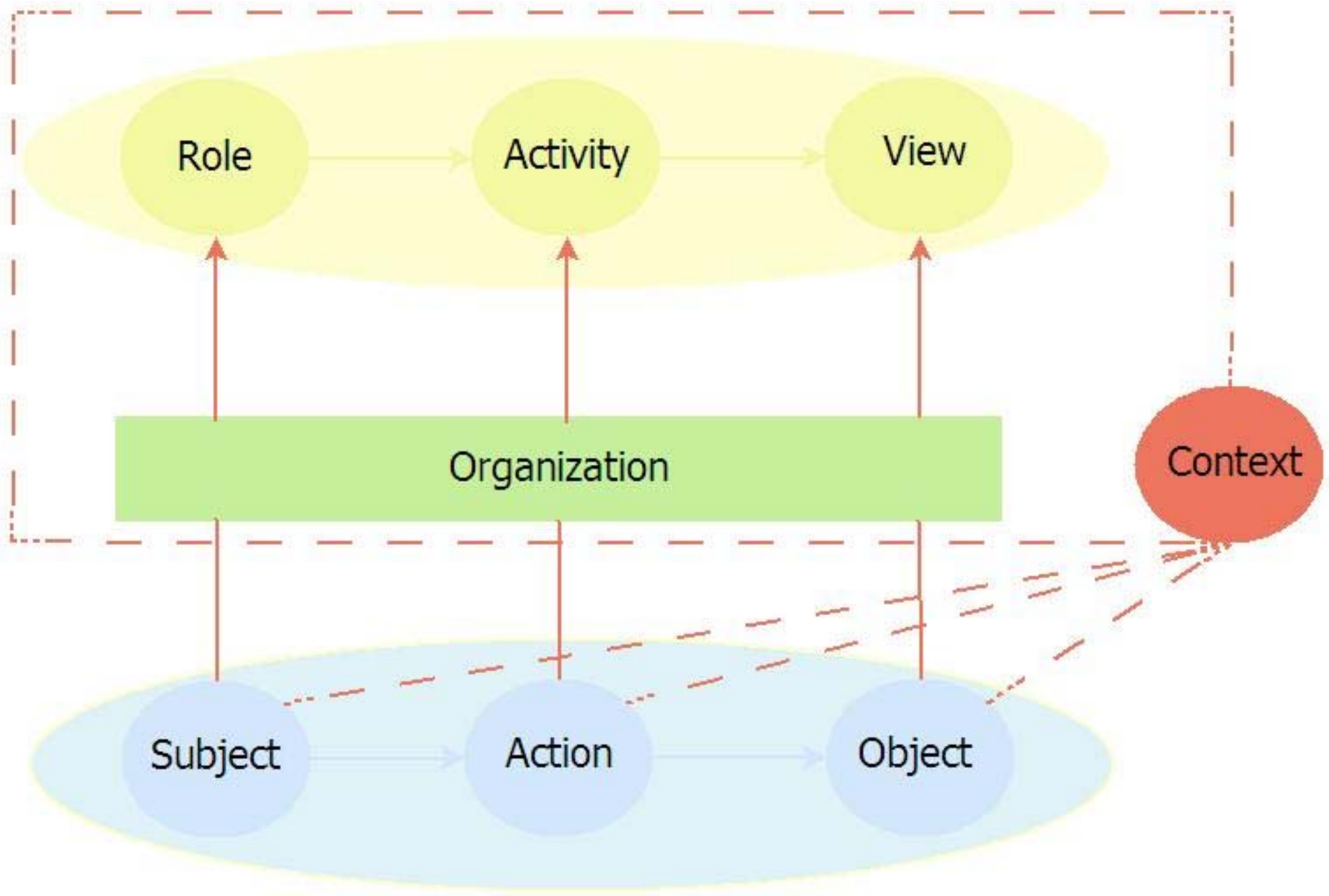}                     %
\caption{The OrBac model}                                                %
\label{The OrBac model}                                                  %
\end{figure}                                                             %

Thus in OrBAC, a role is a set of subjects on which we applied the same security rules. Similarly, an activity is a set of actions which are applied on the same security rules, and the view is a set of objects which are applied on the same security rules.\\
The context is defined for an organization, a subject, an action and the given objects. Furthermore, Contexts can express permissions or prohibitions under certain circumstances (emergency hospital, hours of work in a company, ...). It is easy to imagine that in an emergency context, we desire that a nurse can access the lambda patient records without having to call the administrator, that should gives her the access rights (perhaps too late after). This possibility of varying authorizations is not offered by models like DAC, MAC, RBAC, while in several organizations (hospital, business, ...) there is a real need to grant privileges only in specific circumstances. \\
To conclude, a security rule is defined in OrBAC as follows: P (O, R, A, V, C) where:
\begin{itemize}
 \item P: is a permission, prohibition, or obligation.
 \item O: is an organization.
 \item R: is a role.
 \item A: is an action.
 \item V: is a view.
 \item C: is a context.
\end{itemize}
Example: Permission (Computer Science Department, Administrator, consultant, card customers, hours of work).\\
That is to say, give permission to the administrator of the department to consult the card during office hours.\\
OrBAC model has been implemented in a prototype called MotOrBac made by SERES the team of ENST. MotOrBac allows to specify the modelling and define security policies based on OrBac.

\subsubsection{Security tests}
Establishing mechanisms to protect an application depends on two phases: code securing and modeling the security policy of the application; likewise, verifying the security of an application is made with security tests. Thus, we will apply different kinds of security tests on applications to determine the effectiveness of security mechanisms.\\

Figure 3 shows that the stage of security testing is the final step in securing application. If test results are negative then the application is more or less secure. on the other side, if results are positive we must retry the steps in order to cancel security vulnerabilities that are detected during the phase of security testing. then you must repeat the tests and the steps until do not detect any security breach.
\begin{figure}[h!]                                                       %
\centering                                                               %
\includegraphics[width=8cm,height=3cm]{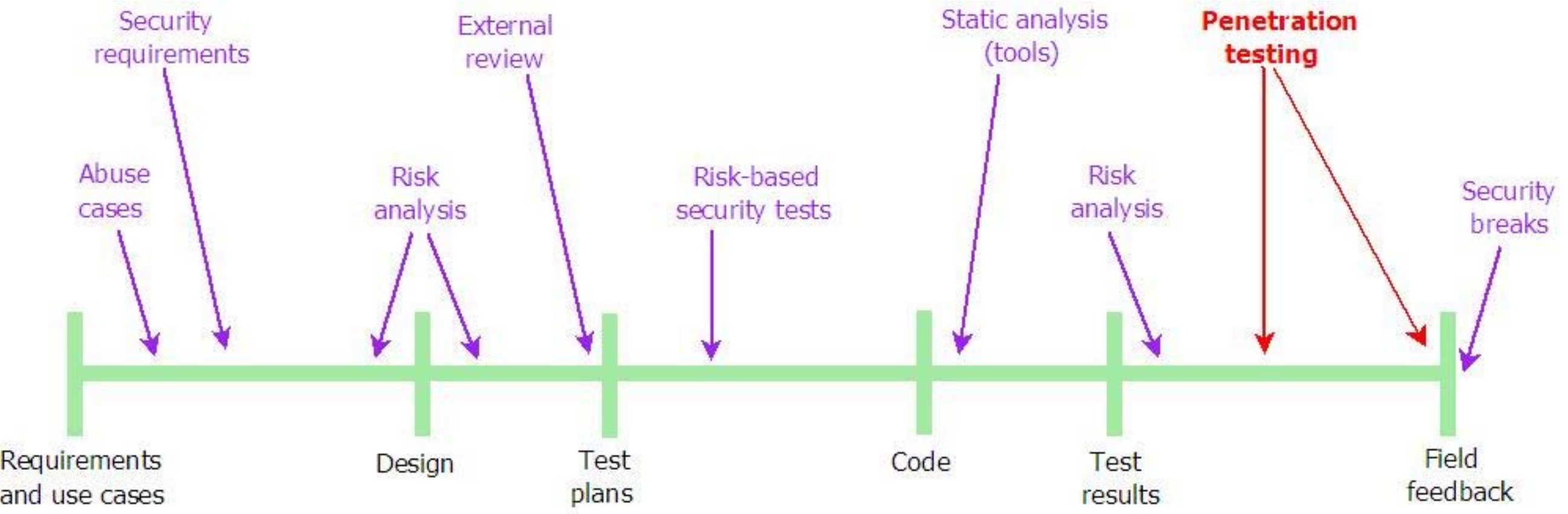}              %
\caption{Security and applications development cycle}                    %
\label{Security and applications development cycle}                      %
\end{figure}                                                             %

\paragraph{Modelling of Attacks}
We can model and describe attacks using UML diagrams \cite{Ray:2005:TAA:1079837.1079984, Hope:2004:MAC:1009229.1009292}, the use case diagrams show the roles (the actors are both the attackers and the victims), while the state transition diagrams describe the steps of realizing an attack, also sequence diagrams complement the model by describing in details and step by step the progress of the attack. Figure 5 describes the website Cross Site Scripting attack (XSS) using sequence diagram.\\
\begin{figure}[h!]                                                       %
\centering                                                               %
\includegraphics[width=7cm,height=3.5cm]{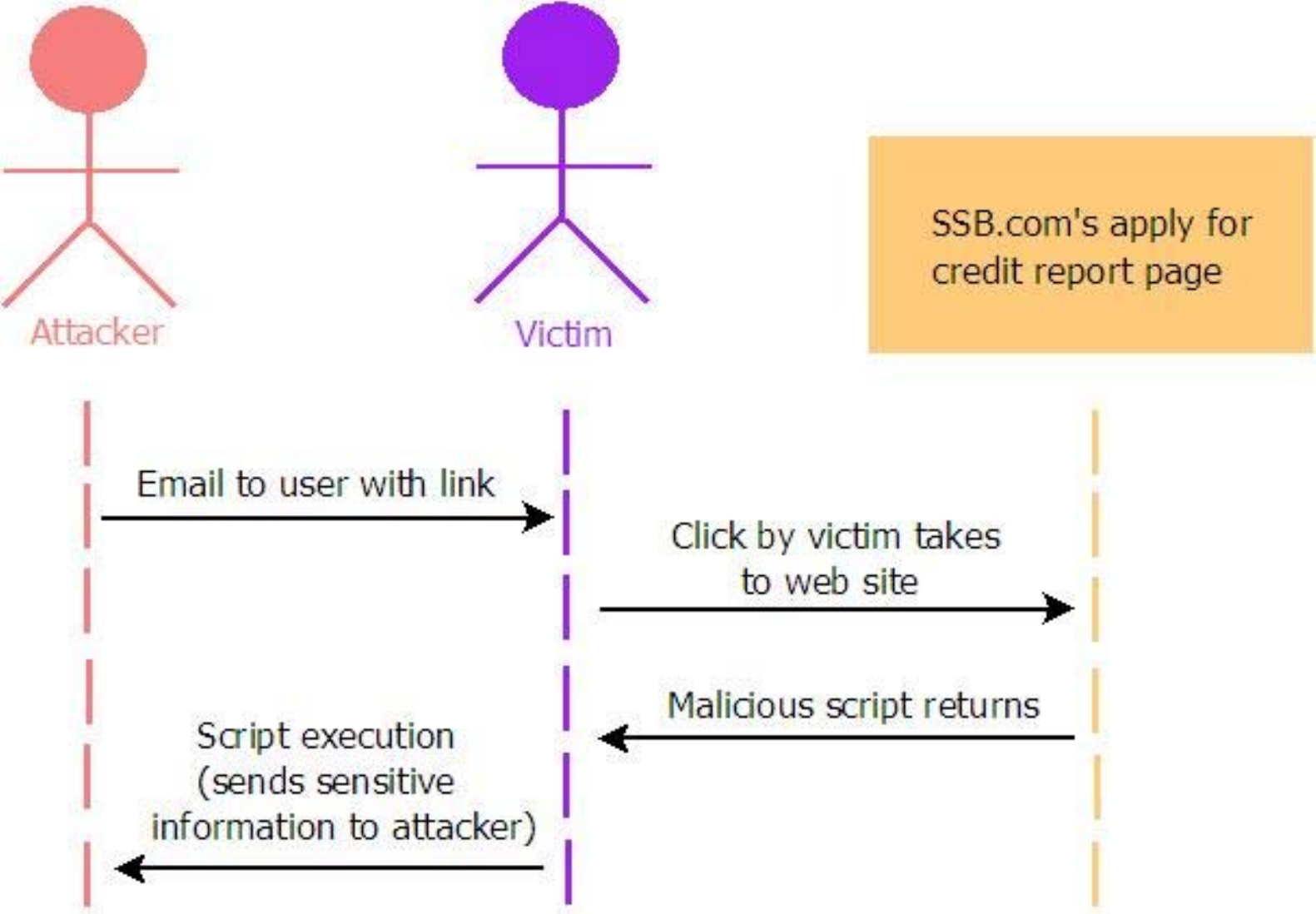}           %
\caption{XSS diagram sequence}                                           %
\label{XSS diagram sequence}                                             %
\end{figure}                                                             %
Modeling attacks in the form of diagrams aims to create a support to penetration testing by teams of special testers (also called Red Team). They model in high-level(rather coarse), the actions to realize to execute these attacks. It should be able to refine these diagrams to consider automation.

\paragraph{Attack simulator}
The role of an attack simulators \cite{Breech:2006:ASS:1190616.1191230} is to automate security tests. Among simulators that exist, we can note the one which offers an original method to test the robustness of security mechanisms, that protect against buffer overflow attack. It is important to know that the simulator does not reproduce the progress of the attack but simulates its consequences on the execution stack; and to accomplish this task, it uses a dynamic compiler that allows it to modify the execution stack. The simulation steps of buffer overflow attack are the following:
\begin{itemize}
 \item Launching the program to test in the dynamic compiler.
 \item The simulator determines if the block of elementary code is vulnerable to attack: the process is to detect if the code uses a predefined risky function, i.e that potentially, code can cause a buffer overflow (copy functions of buffer, unsecured array ...)
 \item The simulator modifies the execution stack to simulate the attack: this step consists on modifying the data stack to create an overflow, as a potential hacker will do.
 \item The processor executes the instructions: The impacts of stack overflow that simulates the attack are propagated in the system, as a real attack.
 \item Observing the reaction of the protection mechanism to determine if the simulation of attack will fail.\\
\end{itemize}
\begin{figure}[h!]                                                       %
\centering                                                               %
\includegraphics[width=7cm,height=3.5cm]{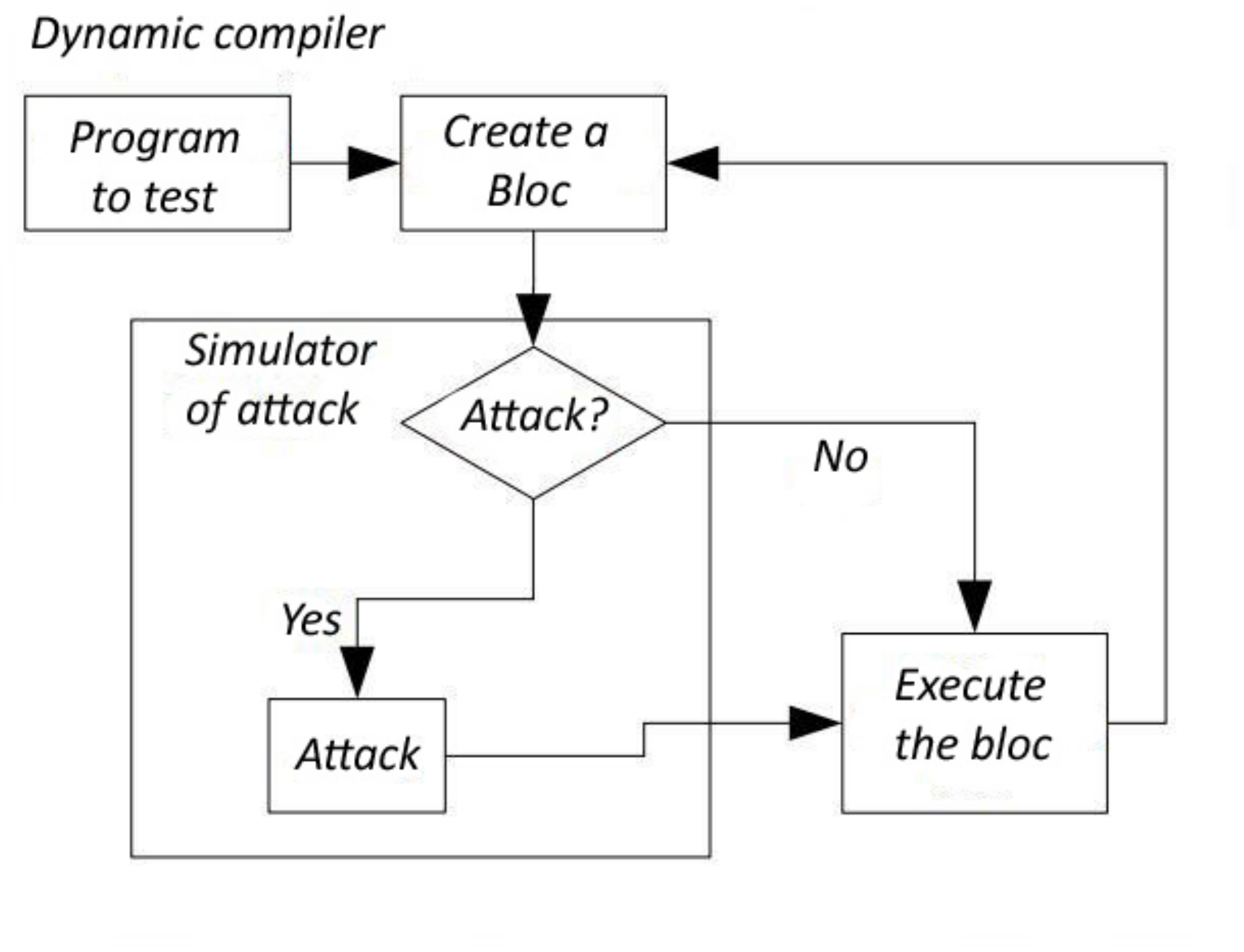}        %
\caption{The working of buffer overflow attack simulator}                %
\label{The working of buffer overflow attack simulator}                  %
\end{figure}                                                             %

This simulation attacks approach is interesting since it's automated, rapid and systematic. It should be noted that it is very recent and still limited to a particular type of buffer overflow attack. Even if, the inventors propose to generalize their approach for other types of buffer overflow attacks as well as the whole of attacks, it still not usable; for example, we cannot simulate the code injection attacks because they are beyond the scope of simulator use.

\paragraph{Penetration tests}
The penetration test \cite{DBLP:journals/ieeesp/ArkinSM05} is an offensive method for evaluating information systems against most famous attacks. In order to apply this type of tests, we will be in need of a team that will play the role of a pirate, and we should follow the steps below:
\begin{itemize}
\item Analyze the system to evaluate, to well understand the conception of his different components and have an idea about the attacks that can be used and the scripts employed.
\item Before applying scripts of attacks, it is necessary to adapt these ones to the systems evaluated. These scripts are programs using the best and well known attacks in the world, that was successful in the past on other system information and which we strongly encouraged to test in this system.
\end{itemize}

The application of penetration tests gives us two results, which we will present in the following:
\begin{itemize}
\item If the script succeeds the attack against information system (since most developers repeat the same security errors), then the system is vulnerable.
\item If the information system succeeds the penetration test, we cannot judge whether the system is secure or not because:
\begin{itemize}
\item It remains vulnerable to new exploits (\textbf{Zero Day Exploits}).
\item Or the attack script is unable to attack the information system and in this case there is a problem of confidence and quality of scripts attacks.
\end{itemize}
\end{itemize}

\textbf{\underline{Problematic:}} The lack of confidence and quality given to the attack scripts of penetration tests.\\
\textbf{\underline{Solution proposed:}} Applying the technique of mutation analysis used in the field of software test.

\subsection{Presentation of Mutation analysis}

The Mutation Test \cite{yve:pas:hel} is a technique that was proposed by De-Millo and it  consists on creating a set of faulty versions of the test program called mutants. The goal for the tester is then to write a series of tests that can distinguish the original program from all its mutants. This technique aims to write relevant tests and involves injecting faults. If the tests applied to the program are able to detect these faults, they can be considered as relevant to disclose faults.\\
The Mutation Analysis is based on the following steps:
\begin{itemize}
 \item Create from a program P, a set of programs Pi called mutants.
 \item Pi should be different from p by one and only one elementary modification introduced in syntactically correct source code of P. This change is called mutation.
 \item Mutations are defined by operators that can be:
 \begin{itemize}
  \item Replacement of an operator by another operator.
  \item Changing a numerical value.
  \item Replacement of a symbol (symbol name of a constant, a variable name, a table, ...).\\
 \end{itemize}
 This yields a set of mutants, each one containing a single error. For example in the table below, we have replaced the less-than symbol ($<$) by the greater-than one ($>$) to have a mutant program. \\
\begin{center}
\begin{tabular}{|l|l|}
\hline Original program & Mutant program \\
\hline void min(int x,int y)\{ & void min(int x,int y)\{\\
               int minval=x; &   int minval=x; \\
        if(y$<$x) minval=y;  &  if(y$>$x) minval=y;\\
        return minval;  &  return minval;\\
       \} & \}\\ \hline
\end{tabular} \vspace{0.3cm} \\
\end{center}
 \item Among the generated mutants, some are \textbf{equivalent} to the original program. That is to say that no input data can distinguish the two programs.\\
On the following table, the program minval is replaced by mutant x in the first statement of the min function. Thus, at this point, the program minval and x will always have the same value, meaning that the mutant program is equivalent to the initial program and both programs give the same results.\\
\begin{center}
\begin{tabular}{|l|l|}
\hline Original program & Mutant program \\
\hline int min(int x,int y)\{ & int min(int x,int y)\{\\
        int minval=x; &   int minval=x; \\
        if(y$<$x) minval=y;  &  if(y$<$minval) minval=y;\\
        return minval;  &  return minval;\\
       \} & \}\\ \hline
\end{tabular} \vspace{0.3cm} \\
\end{center}
 \item The elimination of equivalent mutants is important before proceeding with the analysis, as it is impossible for a test to detect the equivalent mutant.
\end{itemize}
As a next step, we can run tests on the remaining mutants (after eliminating equivalent mutants). If the test performed on a mutant produces an output different from that of the original program, then the test detected the mutant and it is said that the test has \textbf{killed} the mutant. Otherwise the mutant is \textbf{alive}.\\
When a mutant was killed, it can be removed from the testing process because the faults that it contained were detected and this allowed to locate the pertinent test. However, if we keep the mutant in the process, we will run each test on all mutants and we will be able to associate with each test a \textbf{score of a mutation} that corresponds to the proportion of mutants that the process has killed. This score is used as a quality index of a test. If a test T killed m mutants and there are a total of M mutants, the mutation score test T is:
\begin{center}
\fbox{
\begin{minipage}{0.2\textwidth}
\textbf{SM (T) = m / M}
\end{minipage}
}
\end{center}
The overall process of generation test by mutation is therefore consisting on generating all mutants of the program, then creating and executing tests on mutants. While some mutants are still alive, we check if they are not equivalent to the initial program, if they are not, we add tests to try killing them; and we loop until getting a mutation score that satisfy us. This process of mutation analysis is described in Figure 6 :
\begin{figure}[h!]                                                       %
\centering                                                               %
\includegraphics[width=8cm,height=4cm]{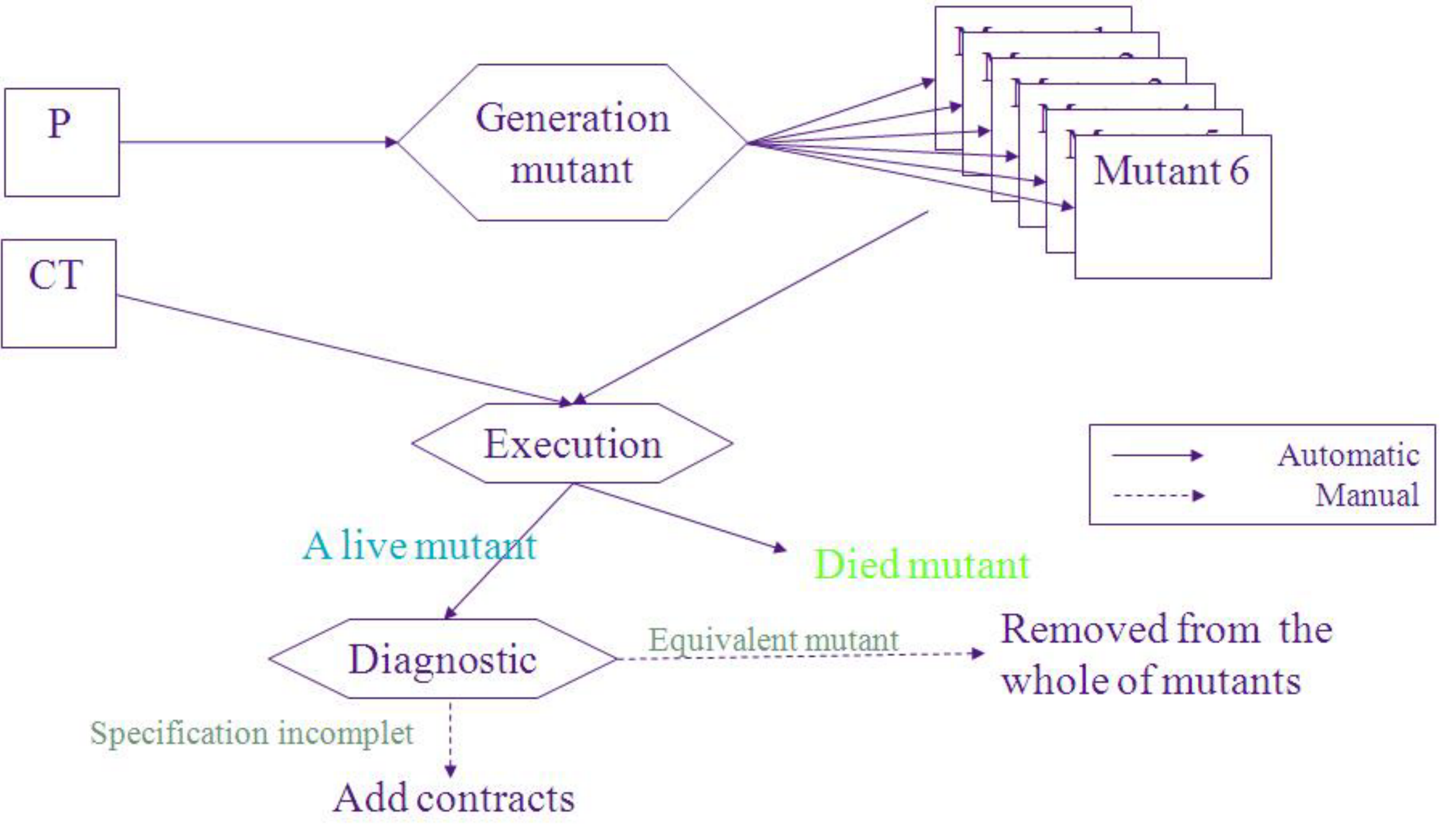}     %
\caption{The working of mutation analysis}                               %
\label{The working of mutation analysis}                                 %
\end{figure}                                                             %
There are several tools of transformation, but Mujava is the most available and most widely used tool, although it is not really effective.

\subsection{Mutation Analysis and Securing applications}
Giving an approach two fields, software tests (Mutation Analysis) and computer security (Penetration Tests) will lead us to a correspondence between their elements. Thus, during this study we will focus on two important points:
\begin{itemize}
\item The first one is the Mutation Operators dedicated to security, therefore, the creation of the security mutants.
\item The second one is the use of these mutants to qualify the Penetration Tests.
\end{itemize}

\subsubsection{The Mutation Operators dedicated to security}
A fault inside a security framework is seen as a security flaw. Therefore, the injection of a fault in the security is only an injection, but at this time, it is a security flaw. This latter is generally based on the perturbation of the application environment using the interaction points of the application such as variables, files and processes.\\
Consequently, a Mutation Operator is a modification or a deletion of an interaction point, on condition that this will create a perturbation in the application environment, and more specifically in the security, which implies the creation a security flaw.\\
The following figure summarizes the procedure of the identification and the creation of the mutation operators.

\newpage                                                                %
\begin{figure}[h!]                                                       %
\centering                                                               %
\includegraphics[width=7.5cm,height=3.5cm]{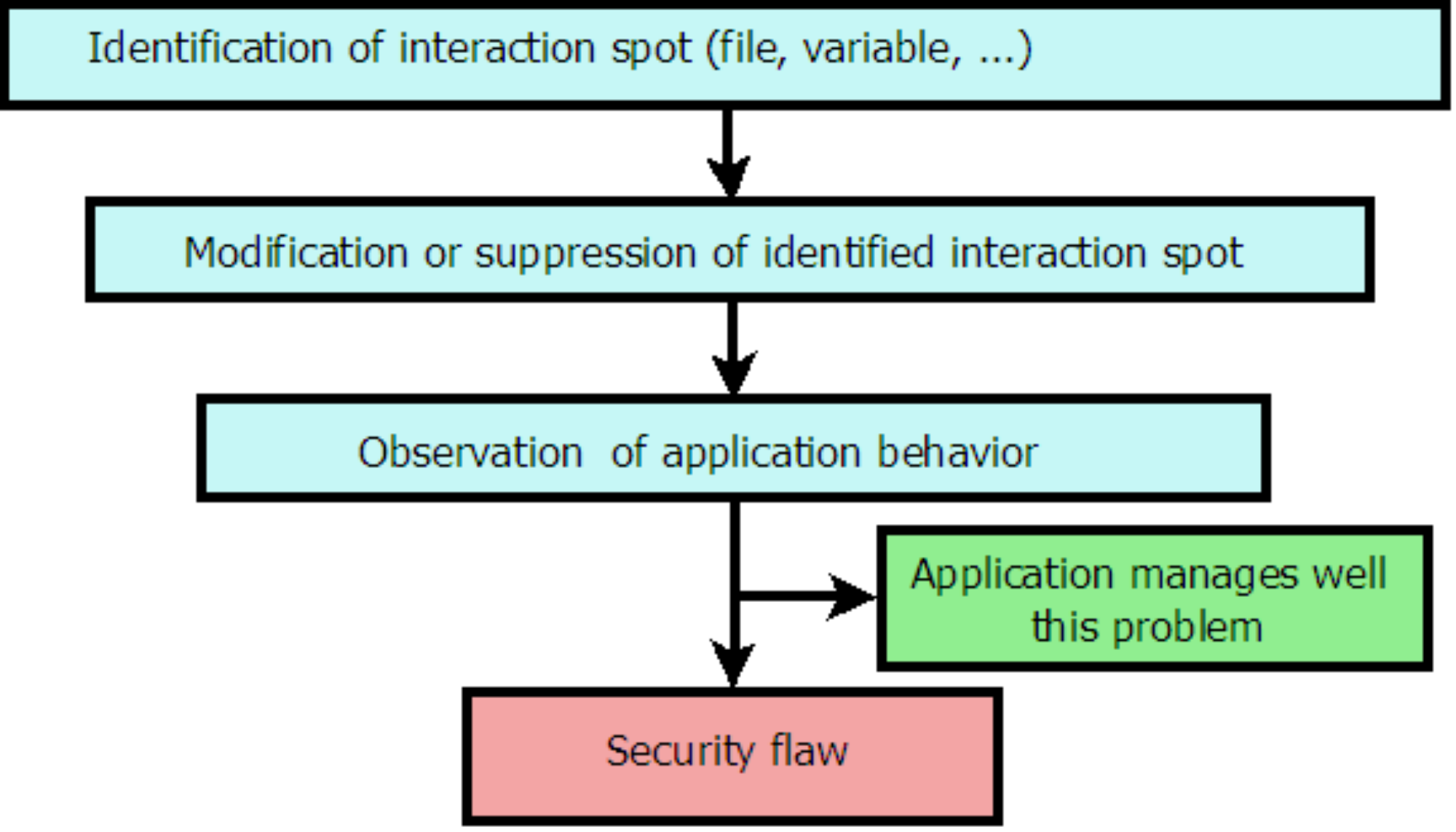}    %
\caption{Flaws injection technique}                                      %
\label{Flaws injection technique}                                        %
\end{figure}                                                             %
The main disadvantage of this technique consists on its difficulty to be automated. Once the point of interaction is perturbed, the conception of the attack scenario can be done only manually.
Among the Mutation Operators that are needed to create mutants for the security tests:
\begin{itemize}
\item The operators that will remove the verification code from user data that are put in dynamic requests, such as authentication code and distribution of access rights.
\item The operators that will eliminate the effect of input control code at the level of communication interfaces between the user and the application.
\end{itemize}

\subsubsection{The qualification of Penetration Tests}

The mutants are the variants of the original application which has released some of its protective mechanisms. A security mutant differs from the reference application by the introduction of errors in one or many of its security mechanisms. Thus, we can couple an error in the input user treatment with an error in the implementation of the security policy. We are therefore looking for creating mutants by mutating the control code of input. Moreover, we can couple a flaw in the security interfaces with vulnerability introduced voluntarily in the security policy (for example, in access to the database, we can give a write access to an application that has only read access).\\
To illustrate this approach, the figure 8 shows a simplified points in which we can voluntarily introduce a flaw in an information system connected to the Internet, such as: the communication interfaces with users, management of buffers, stacks of execution and implementation of the security policy (for example, defined with the model Or-Bac). We take as a prototype of test, a client-server application 3-tier type (so with database). This database will implement security policies (created from MotOrBAC for example).

\begin{figure}[h!]                                                       %
\centering                                                               %
\includegraphics[width=7.5cm,height=4cm]{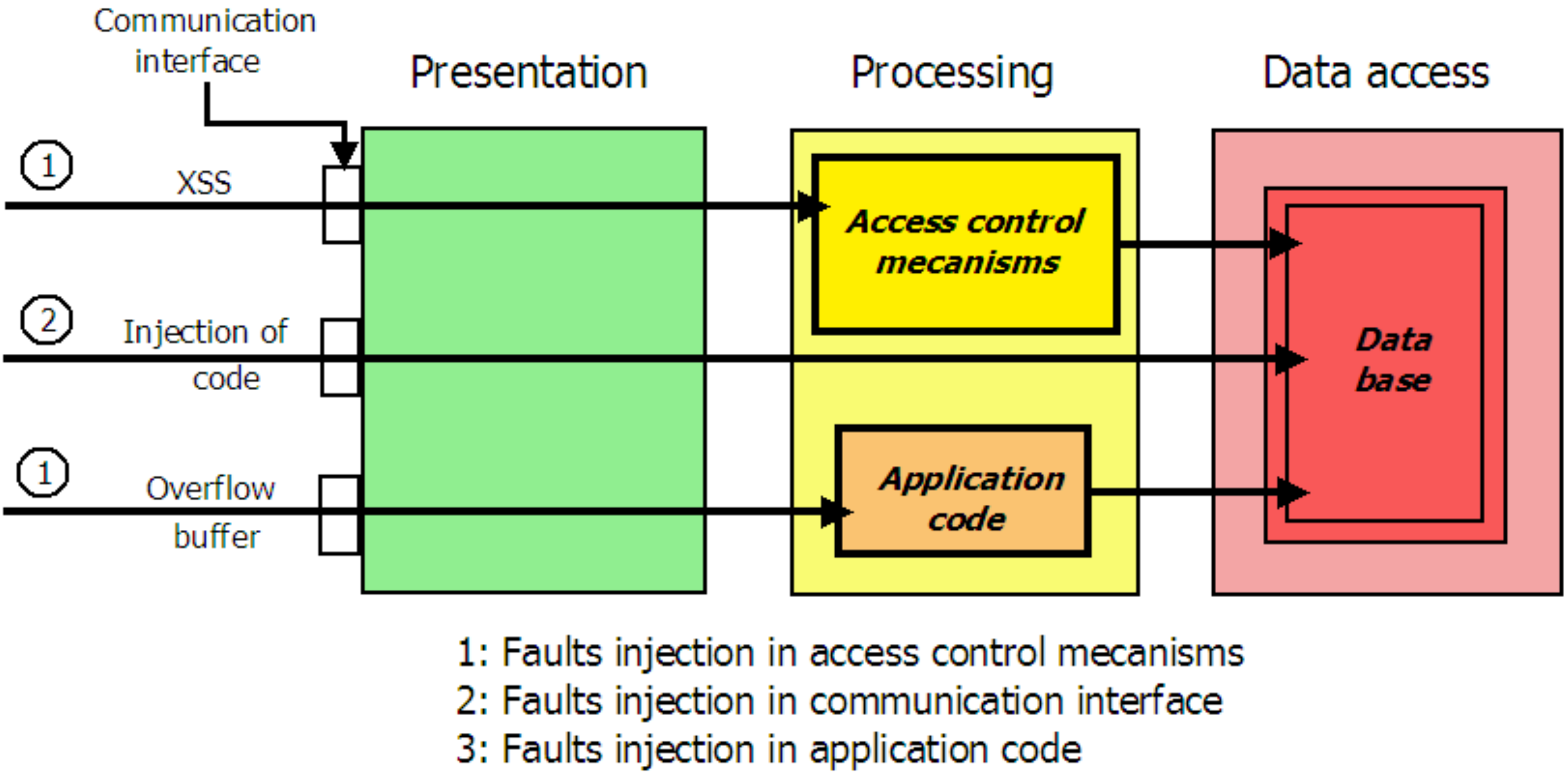}               %
\caption{Mutant model of testing attacks in 3-tier application}          %
\label{Mutant model of testing attacks in 3-tier application}            %
\end{figure}                                                             %
There are similarities between the software test and the penetration test. Both approaches aim to find bugs and errors in applications. However, there are some points of divergence. Bugs are easier to find, in view of that the application produces erroneous execution traces. Yet, security flaws are generally more subtle to detect. Indeed, the application having security flaws can behave in a normal way concerning features it ensures, this is the case of flaws related to the injection of code: The interpreted application code is correct, but it is vulnerable to attacks. So the attack scripts that will test cases must not only pass the level presentation, but also access the database (even modify it).\\
For the evaluation of the test results of attack, the execution traces cannot be used (except for the stack overflow attacks when we get a segmentation error). This implies that the success of the attack depends on the access of test script to the database, in which it attempts to execute an operation, meaning that the success of the test depends on the success of this operation.\\
We can summarize our approach in the following way:\\
The elements:
\begin{itemize}
\item An application that will be used to evaluate the security tests.
\item The script of penetration test that is the test subject.
\end{itemize}
The steps:
\begin{enumerate}
\item The creation of mutants:
\begin{enumerate}
\item The identification of the interaction points in the application that will create security flaws. Therefore, the identification of mutation operators.
\item The creation of security mutants by using at least one mutation operator.
\end{enumerate}
\item The evaluation of the penetration test:
\begin{enumerate}
\item Applying the attack script of this penetration test to the mutants created.
\item Observing the effect of the evaluation:
\begin{itemize}
\item If the attack script succeeds to do access of at least one mutant to the database. So, the penetration test corresponding to this script is considered a well qualified test .
\item Else, the penetration test is disqualified.
\end{itemize}
\end{enumerate}
\end{enumerate}

\section{Mutation Analysis and Access\\ Control}
\subsection{Presentation of Access Control}
\subsubsection{The access control}
Access Control is the data protection against threats of confidentiality (disclosure unauthorized), integrity (Unauthorized modification)and availability (denial of service) when sharing data between multiple users.
To ensure this protection, each data access must be controlled and obviously all unauthorized access must be strictly blocked.\\
The development of an Access Control model is based on the definition of access control policies that determine access permissions to data. This model ensures that data are accessible only for users having the access rights. In gross, Access Control ensures that user access to resources must respect a well-defined security policy.\\
The establishment of a general view, which is not limited to one existed access control model (DAC, MAC, RBAC, Gold-Bac, ...) will lead us to use a model for hierarchical architecture with two types of security requirements (permissions and prohibitions), and which incorporates the principle of constraints (temporary, Space, ...) that has as name "the context".\\
A rule of our model is described as follows \textbf{P(R, A, C)} where: \textbf{P:} permission or prohibition; \textbf{R:} role; \textbf{A:} action; \textbf{C:} context.\\
Take the example of a faculty management system that offers management features of students, teachers and staff.
In this application, the user can perform two operations: view their account and access to the courses. The resources that will be controlled are the accounts, marks and folders of users and staff. Entities A and R can be ordered hierarchically. For example in the management application of a faculty, there are two types of roles: user and personal. Concerning student and teacher roles, they inherit the user role. This hierarchy allows to define rules at the user's role level, to be then applied to the student and teacher roles. Finally, there are five temporal contexts: the work duration, the period of deliberation, studies, holiday and default, all this entities are detailed in Figure 9. Once the entities defined, the access control policy must be written. Examples of rules:
\begin{itemize}
 \item[\textbf{R1:}] Permission(Administrator, create account, work duration)
 \item[\textbf{R2:}] Permission(user, user activities, work duration)
 \item[\textbf{R3:}] Prohibition (Teacher, enter marks, studies)
\end{itemize}
There are two types of rules: primary and concrete. The primary rules are the set of defined ones. By cons, concrete rules are all those obtained after deriving primary rules basing on the hierarchy (which is obtained by the relation: specification / generalization), it is the case of R2 rule, which is applied to "user activity" and designates all the actions that user can make (which are from Figure 9: Consult his account, Benefit from services, Access to courses).
\begin{figure}[h!]                                                       %
\centering                                                               %
\includegraphics[width=9cm,height=5.5cm]{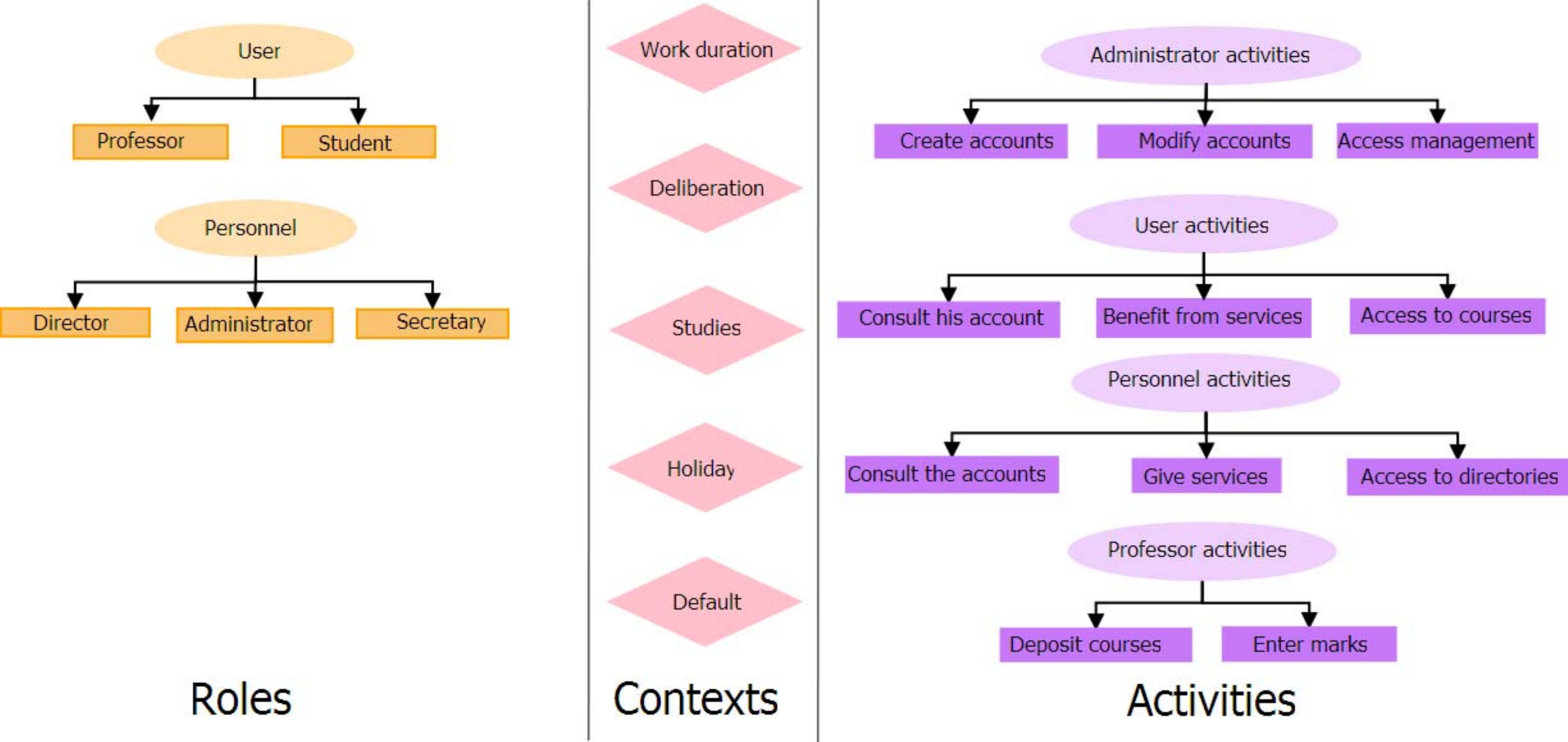} %
\caption{The Entities of a management system of a faculty}               %
\label{The Entities of a management system of a faculty}                 %
\end{figure}                                                             %
We must distinguish between  primary  and concrete rules. If an actor plays mostly the role of rule R2, he simultaneously plays the role of the rule R1. Therefore, we say that the rule R1 is primary and R2 is concrete. For Example in the hierarchy shown in the diagram above, the teacher is also a user. So, whenever a person is associated with the role of teacher, he also plays the role of the user.

\subsubsection{Integration of an access control policy Architecture in application}
An application implementing an access control policy must have an architecture with two separate parts, the first named Business Logic (BL) contains the code that implants functional requirements of system, and the second called Access Control (AC). For instance, the management system of a faculty contains code to enter marks, modify accounts and update the database.\\
The access control policy is encapsulated in a component called PDP (Policy Decision Point), which can store the access control policy either in a text file or a database, to implement security rules and then set or change the policy. The PDP is used by PEPs (Policy Enforcement Point) placed in points of application where they will execute the security rules, just before the execution of features that are monitored by the access control policy. PEPs have a dual role, first of all, they send requests to the PDP with information about the user's role, his activity, the view demanded and the contexts. The PDP then decides if the requested access is permitted or prohibited, basing on the rules of the access control policy and sends this decision to the PEP concerned that will apply it.\\
The PDP is based on rules of an access control policy and as this latter is generated automatically, it is also automatically generated. By cons, the PEP operates in a manual way to adjust successfully to the code and application architecture.

\subsection{The tests and mutation operators designed for access control}
\subsubsection{Mutation operators designed for Access Control}
The fact that the business logic party is separated from the access control policy, will simplify for us the injection of faults in the access control mechanisms. So, we proceed to the replacement of used policy by the PDP, to get an implementation of the mutant policy.\\
The creation of mutants is performed systematically via mutation operators, where each one injects a particular type of error and only one error is injected at the same time to create each mutant. With \cite{Tej}, we proposed the following mutation operators:
\begin{center}
\begin{tabular}{|p{1cm}|p{7cm}|}
  \hline
  \multicolumn{2}{|c|}{Basic operators changing the type} \\\hline
  PPR & (permission to prohibition) replaces a rule of permission by a prohibition.\\ \hline
  PRP & (prohibition to permission) replaces a rule prohibiting by permission.\\ \hline

  \multicolumn{2}{|c|}{Basic operators changing parameters} \\ \hline
  RRD & (role replaced with different one) replaces the role of a rule by another role selected randomly.\\ \hline
  CRD & (context replaced with different one) replaces the context of rule by another context selected randomly.\\ \hline

  \multicolumn{2}{|c|}{Basic operators modifying the hierarchy} \\ \hline
  RPD & (parent role replaced with a descendant) replaces a role of a rule by one of its descendants (thus changing the derived rules)\\ \hline
  APD & (Parent Action replaced with a descendant) replaces an action in rule by one of its descendants (thus changing the derived rules) \\ \hline
\end{tabular}
\end{center}

\begin{center}
\begin{tabular}{|p{1cm}|p{7cm}|}
  \hline
  \multicolumn{2}{|c|}{Advanced Operator} \\ \hline
  ANR & adds a new rule not belonging to defined rules part.\\ \hline
\end{tabular}
\end{center}
It is important to emphasize that these operations do not produce equivalent mutants. Indeed, the mutants created by construction, contain different rules  from the initial policy because they add a new rule or modify an existing one.\\
There are two types of mutation operators, the basic operators (all operators except ANR) and the advanced mutation operators called ANR, that are special because they aim to test the default behavior of access control mechanism. Truly, any policy of access control contains X rules and a default one (permission or prohibition) that is applied when the other rules do not correspond to entities of a query. The ANR operator can create rules that replace the default rule and complement the X defined rules. That's why it's an advanced operator as it allows to test the robustness of the access control mechanism.

\subsubsection{The creation of access control tests}
There are two types of tests \cite{Tej}: functional and security ones. The first type aims to test the functionality of an information system, while the second type is designed to test the security of an information system. Concerning security tests generated from the access control policy, they aim to validate the compliance of the security mechanism with its access control policy. Both types are interconnected as the functional tests also validate compliance with the access control policy.\\
Figure 10 shows the three parts of a test, whether functional or security:
\begin{figure}[h!]                                                       %
\centering                                                               %
\includegraphics[width=8cm,height=2cm]{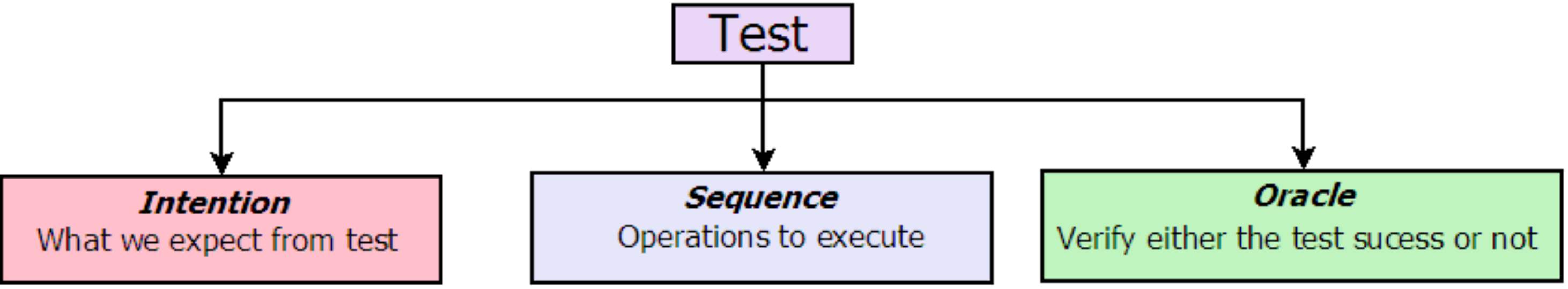}        %
\caption{The three parts of a test}                                      %
\label{The three parts of a test}                                        %
\end{figure}                                                             %
\\
To illustrate the difference between functional and security tests, we present two simple examples:
\begin{enumerate}
 \item Functional test: Tests student accesses to a course:
 \begin{itemize}
  \item \textbf{Intention:} Test whether downloading a course is available for the student.
  \item \textbf{Sequence:} Enter the account and download a course while in school.
  \item \textbf{Oracle:} The number increases during download.
 \end{itemize}
 \item Security test: Tests the ability of a student to access a course during studies (as specified by CA policy):
 \begin{itemize}
  \item \textbf{Intention:} Tests whether a student has the right to download a course in the worked days .
  \item \textbf{Sequence:} Enters the account and download a course during studies.
  \item \textbf{Oracle:} Questioning the PDP to ensure that the right rules are enabled.
 \end{itemize}
\end{enumerate}
The Generation of security tests is based on several criteria representing the test objectives to evaluate the information system and decide whether a problem exists or not.\\
We propose the following three criteria:
\begin{enumerate}
 \item \textbf{All primary rules:} Cover the primary rules, if the rule is differentiable, then test one of its derived rules.
 \item \textbf{All the concrete rules:} Cover the entire concrete rules.
 \item \textbf{All default rules:} This criterion is intended to validate all the default rules not specified in the security policy.
\end{enumerate}
\begin{figure}[h!]                                                       %
\centering                                                               %
\includegraphics[width=8cm,height=3cm]{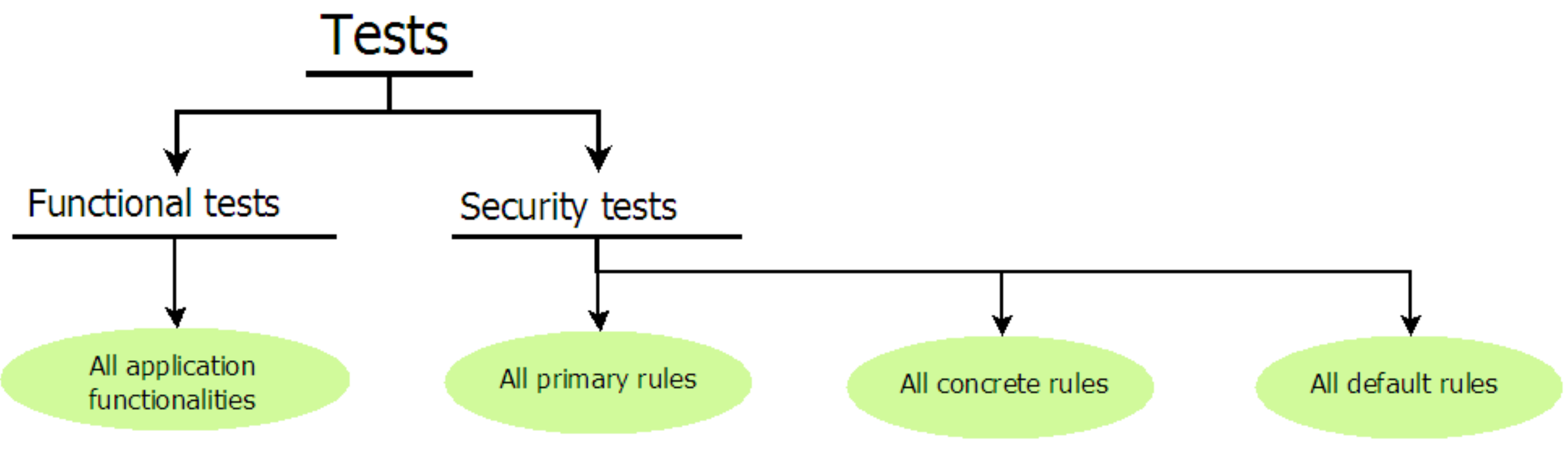}          %
\caption{The different types of tests and their criterias}               %
\label{The different types of tests and their criterias}                 %
\end{figure}                                                             %
The definition of a primary rule accords to its differentiability or not, we can say that it's also concrete if it derives from itself. By cons, a concrete rule is a rule that inherits from an another one.\\
With this logic, and using only the second criterion (all concrete rules), tests can cover all the rules of the access control policy, whether primary or concrete. That leaves, to cover the default rules and that is the objective of tests generated from the third criterion (all default rules).\\
To conclude, we must say that the right strategy is to use very specific security tests and to combine the two criteria covering both: all concrete and all default rules.

\subsection{Mutation Analysis and hidden mechanisms for access control }
The best approach to implement an access control policy is to separate PDP from the  business logic part. However, separation is rarely perfectly clear, especially in older systems. When it comes to change the access control policy of these applications, it is necessary to locate these places in the code that prevents the system evolution, because they implement directly the access control rules in the code.\\
In this section, we will outline the use of mutation as an effective way to detect hidden mechanisms. We will first describe the problem encountered which we are trying to solve and then present the approach based on the mutation. We will also present a general mutation approach to guide the evolution of system access control policy .

\subsubsection{The hidden mechanisms of access control}
The access control mechanisms can be of different natures. Thus, there are explicit visible mechanisms, and implicit hidden ones:
\begin{itemize}
 \item The explicit mechanisms are implanted in the code:
 \begin{itemize}
  \item The explicit mechanisms are visible when you have a traceability link or documentation to establish a relationship with the access control policy (or part thereof). For example, in Figure 12 the visible explicit mechanism uses an external component (which is the "security Policy Service") responsible for applying security policy, this controllable component is based on an access control policy .
  \item An explicit mechanism is said hidden when we don't have any information to link this mechanism with a rule in the access control policy. For instance, in Figure 12 the hidden explicit mechanism directly applies an access control rule (denying access based on a condition).
 \end{itemize}
 \begin{figure}[h!]                                                       %
 \centering                                                               %
\includegraphics[width=8.5cm,height=4cm]{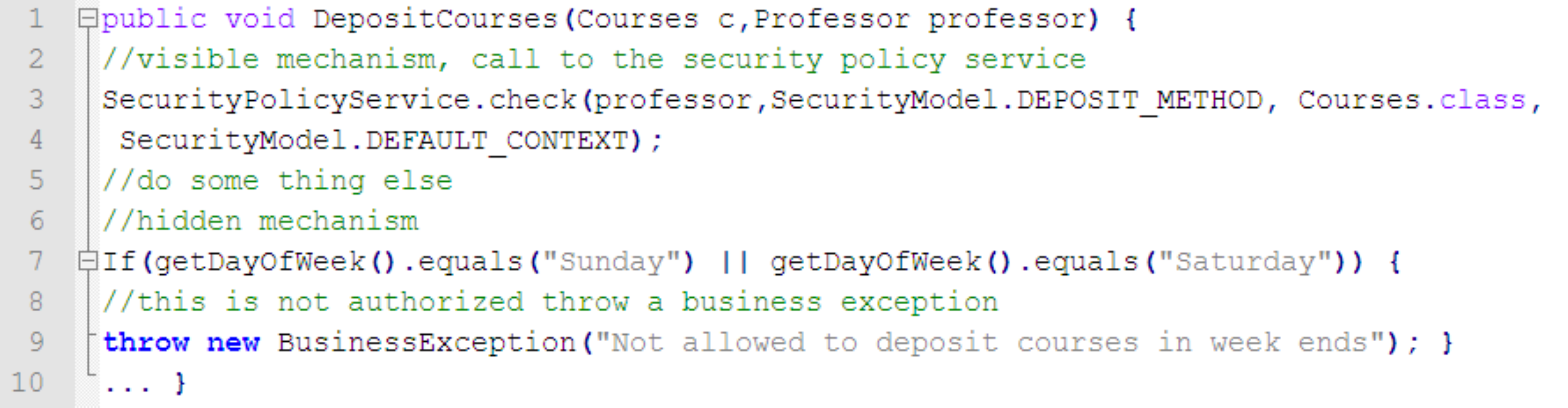}       %
 \caption{Example of explicit access control mechanisms}                  %
 \label{Example of explicit mechanisms for access control}                %
 \end{figure}                                                             %
 \item The implicit mechanisms represent access control constraints integrated into the architecture itself of the code or system. The figure 13 illustrates an example of an implicit mechanism, where by construction, the class "Secretary" does not have a method to modify an account. Note that the implicit implantation of an access control rule prohibits access to the secretaries to modify accounts. The access control policy is expressed implicitly via the "class" model, which makes its development more difficult. In this example, the fact of allowing secretaries to change the accounts is due to adding in the class "Server" a new method "Modify Account (secretary, account)". In this case modification is easy, but in other cases, the modification can be difficult and tedious even not impossible (especially when there is a significant coupling between classes).\\

\begin{figure}[h!]                                                       %
\centering                                                               %
\includegraphics[width=9cm,height=4.5cm]{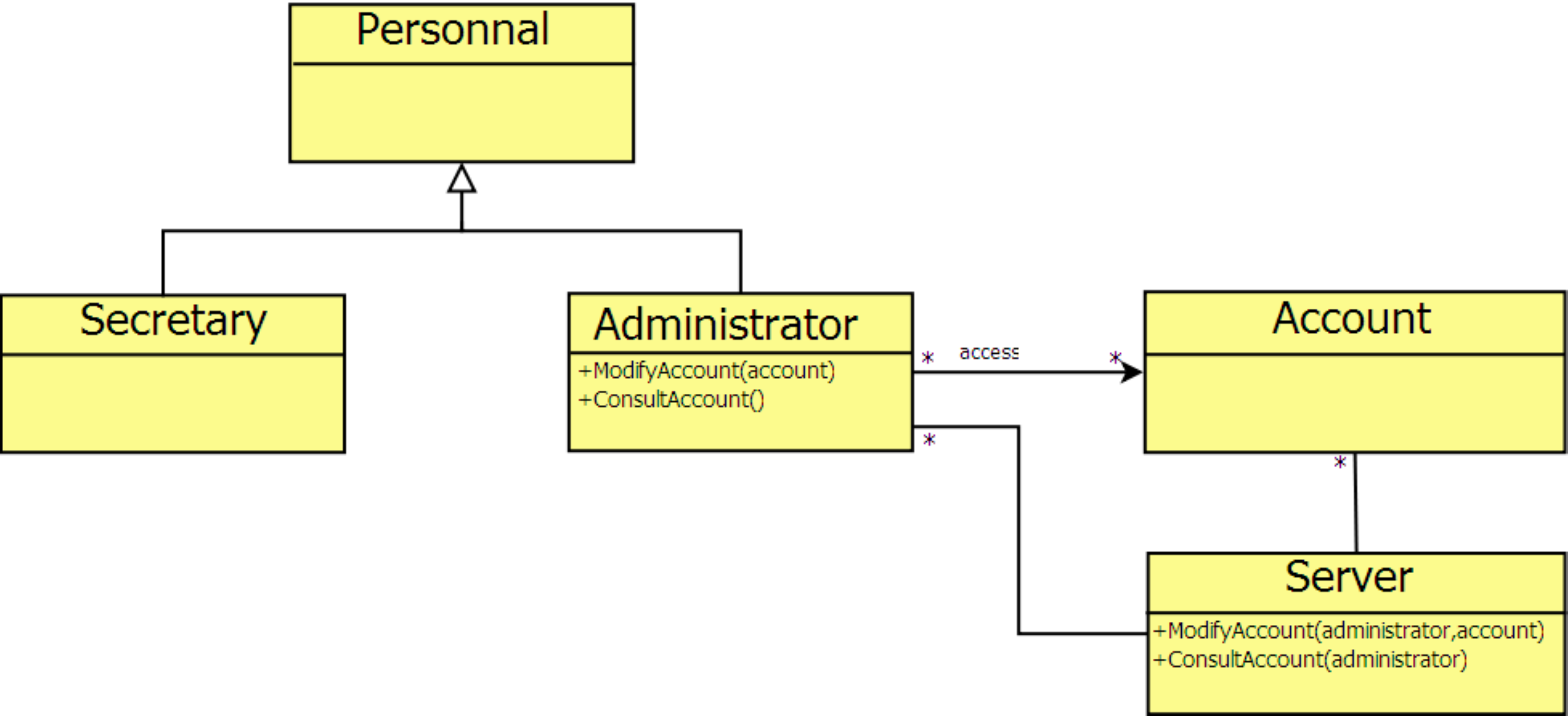}              %
\caption{Example of implicit access control mechanisms}                  %
\label{Example of implicit access control mechanisms}                    %
\end{figure}                                                             %

Modifying an access control rule implicitly implemented by one of these mechanisms, involves doing the "refactoring" (rework)of architecture.
\end{itemize}

The access control mechanisms whether explicit or implicit, must be taken into account during the evolution of the CA policy by following the steps noted below:
\begin{enumerate}
 \item Explicit and visible mechanisms must be modified and tested.
 \item Hidden and explicit mechanisms may be in conflict with the new policy. So, They must be located and removed.
 \item Implicit mechanisms may prevent the development of policy. In this case the model must be modified or redesigned to make the application more flexible.
\end{enumerate}
To Summarize, we can say that there are two hidden mechanisms for access control: explicit hidden mechanisms and implicit ones. These two mechanisms prevent the evolution of the access control policy. Therefore, detect and eliminate these hidden mechanisms will be the purpose of our method based on Mutation Analysis.

\subsubsection{Detection of hidden mechanisms for access control via Mutation Analysis}
We will present an approach that follows the detection method of hidden access control mechanisms. We should respect the order of these steps without forgetting any step:
\begin{enumerate}
 \item Applying the mutation on the access control policy using the initial two change operators of types PRP and PPR, which simulates an incremental evolution of the policy.
 \item Implementing and creating tests from the mutant policy. In other words, the tests created will cover the rules of the access control mutants.
 \item Disabling all visible security mechanisms on the initial information system (ie: we want to evaluate its access control policy ), which is easy because the activated rules are known.
 \item After disabling security mechanisms, the information system doesn't contain any prohibition. So, when launching tests on this system, if these tests are successful, there will be no hidden security mechanism, otherwise, the failed tests necessarily indicate the existence of hidden security mechanisms.
\end{enumerate}
This approach gives us a solution based on tests to detect the hidden access control mechanisms and also the rules subsequently implanted by them. This method allows to find an indicator that estimates the flexibility of an information system. This flexibility is the number of visible rules divided by the total number of security rules (which is visible or hidden).

\subsubsection{Evolution of Security Policy}
A micro-evolution as the name suggests, is a small change in the access control policy. There are two types of micro-evolution:
\begin{itemize}
 \item \textbf{$\delta^{+}$:} which relaxes the policy by adding a permission or removing a prohibition.
 \item \textbf{$\delta^{-}$:} which restricts the policy by adding a prohibition or removing a permission.
\end{itemize}
A security policy is a set of permission and prohibition. Therefore, it is indeed a set of micro-evolution. As shown in the Figure 14 diagram, the evolution of an initial security policy to another final one is simply an application of a set of micro-evolution on the first one to reach a new security policy.
\newpage                                                                   %
\begin{figure}[h!]                                                          %
\centering                                                                  %
\includegraphics[width=4cm,height=0.7cm]{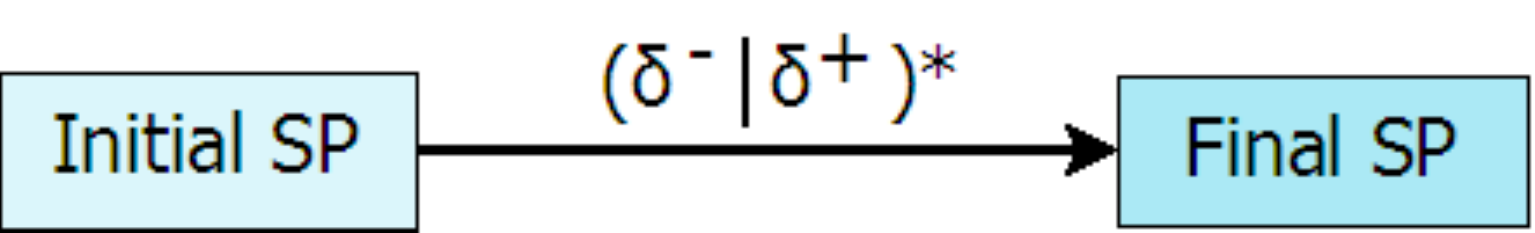}%
\caption{The micro-evolutions}                                              %
\label{The micro-evolutions}                                                %
\end{figure}                                                                %

The following formula shows the evolution of an initial access control policy SPInit to a new access control policy SPNew:\\

\fbox{
\begin{minipage}{0.42\textwidth}
$ \Delta(SPInit) = \delta^{+-}_{n} \circ \delta^{+-}_{n-1} \circ \delta^{+-}_{n-2} \circ ..... \circ \delta^{+-}_{1}(SPInit) \\ \hspace*{1.7cm} = SPNew$
\end{minipage}
}\\

As was previously mentioned, the evolution of the system is a set of micro-evolution. If we can detect the hidden mechanisms that prevent the establishment of each micro-evolution, then we can modify the code to eliminate these hidden mechanisms, meanings that the access control policy can evolve using the set of micro-evolution evaluated.\\
The approach adopted to modify the access control policy of an information system is:
\begin{enumerate}
 \item Formulate new rules of access control policy in the form of a set of micro-evolution.
 \item Create a single mutant from each micro-evolution.
 \item Generate test cases for each mutant policy (ie: each mutant).
 \item Disable visible access control mechanisms in the mutant that corresponds to the micro-evolution.
 \item Start the test generating mutant policy.
 \item There are two possible cases:
 \begin{itemize}
  \item If the test passes. So, no hidden mechanism against micro-evolution.
  \item Otherwise, the existence of hidden mechanisms.
 \end{itemize}
 \item In the second case, we must eliminate the hidden mechanisms since they are detected. Then repeat the method from beginning to end.
 \item Implant and document the micro-evolution in the visible mechanism.
\end{enumerate}

In this part of our study we have tried to give a detailed overview of the access control policies. Then, we have introduced the technique of Mutation Analysis specified in access control and used as a support to evaluate the criteria for generating tests.\\
On the other hand, to change the access control policy of an application, we have proposed a comprehensive approach based on tests and mutation to solve the problem of hidden access control mechanisms.

\section{Conclusion}

This article presents a summary of a study conducted in the field of computer security, specifically, web application security. This study will lead us to make a link between two aspects relating two research disciplines that are: Testing out different software and Security.\\
During the process of collecting and research, we started working on security as a means for securing applications, which allowed us at the first time to see the different criteria and security aspects such as the three levels (application, system and network) and the five main objectives (integrity, confidentiality, availability, no-repudiation and authentication). \\
Thus, after a research conducted on the majority of existing attacks types so far, we are confident that the applications need a security that affects several points, starting with the code, through the modeling of security policies, and arriving to security tests that evaluate the security mechanisms implemented in applications.\\
In this regard, we thought to apply Mutation Analysis as a method of software test field on  scripts of security tests, and that to qualify and ensure their ability to be a true evaluation tool that detects remained flaws in an application after the securing phase.\\
In the second time, we decided to apply the technique of Mutation Analysis on other security disciplines such as access control.\\
Access control tests are similar to other security tests, unless they have a specification to be based on rules established in the access control policy.\\
Thanks to this specification we succeed to precise the exact types of criteria that can generate good quality of access control tests. Then, we have solved the problem of hidden access control mechanisms preventing the development of security policy, by proposing an approach applied during the phase of the evolution of access control policy.\\
In this study, it remains to implement these techniques and to apply them on real cases to reach the experimental results with statistics, to be able to generalize our solution.\\
As perspective, we will try to take the security from different points of view, to migrate the techniques and solutions that are successful in other fields, also to adapt them to the computer security field.

\bibliographystyle{abbrv}
\bibliography{biblio}  
%
%

\end{document}